\documentclass[useAMS,usegraphicx]{mn2e}
\usepackage{amsmath}
\usepackage{color}    					
\usepackage{graphicx}
\usepackage[normalem]{ulem}  			

\title{The mass-metallicity relation of tidal dwarf galaxies}

\author[Recchi et al.]{S.
  Recchi$^{1}$\thanks{simone.recchi@univie.ac.at}, P.
  Kroupa$^{2}$\thanks{pavel@astro.uni-bonn.de} and S.
  Ploeckinger$^{{\bf1,}3}$\thanks{ploeckinger@strw.leidenuniv.nl}\\
$^{1}$Department of Astrophysics, Vienna University,
  T\"urkenschanzstrasse 17, A-1180, Vienna, Austria \\
  $^{2}$Helmholtz-Institut f\"ur Strahlen- und Kernphysik (HISKP),
  Universit\"at Bonn, Rheinische Friedrich-Wilhelms-Universit\"at,\\
  Nussallee 14-16, D-53115 Bonn, Germany\\
$^3$Leiden Observatory, Leiden University, PO Box 9513, 2300 RA Leiden, The Netherlands
 }

\date{Received; accepted}

\begin{document}
\maketitle



\begin{abstract} {  {  Dwarf galaxies generally follow a
      mass-metallicity (MZ) relation, where more massive objects
      retain a larger fraction of heavy elements. Young tidal dwarf
      galaxies (TDGs), born in the tidal tails produced by interacting
      gas-rich galaxies, {  have been thought to} not follow
      the MZ relation, because they inherit the metallicity of the
      more massive parent galaxies.  We present chemical evolution
      models to investigate if TDGs that formed at very high
      redshifts, where the metallicity of their parent galaxy was very
      low, can produce the observed MZ relation. Assuming that galaxy
      interactions were more frequent in the denser high-redshift
      universe,} TDGs could constitute an important contribution to
    the dwarf galaxy population.
    The {  survey} of chemical evolution models of TDGs
    presented here captures for the first time an initial mass
    function (IMF) of stars that is dependent on both the star
    formation rate and the gas metallicity {  via the
      integrated galactic IMF (IGIMF) theory}.  As TDGs form in the
    tidal debris of interacting galaxies, the pre-enrichment of the
    gas, an underlying pre-existing stellar population, infall, and
    mass dependent outflows are considered.  The models of young TDGs
    that are created in strongly pre-enriched tidal arms with a
    pre-existing stellar population can explain the measured abundance
    ratios of observed TDGs. The same chemical evolution models for
    TDGs, that form out of gas with initially very low metallicity,
    naturally build up {  the observed} MZ relation.  The
    modelled chemical composition of ancient TDGs is therefore
    consistent with the observed MZ relation of satellite galaxies. }
\end{abstract}


\maketitle


\section{Introduction}

{  Dark-matter-free tidal dwarf galaxies (TDGs) may constitute an
  important contribution to the dwarf galaxy population.  TDGs are
  dwarf galaxies that form from the tidal debris of baryonic material
  liberated from giant galaxies after they interact with other
  galaxies} ({  see Bournaud 2010 and} Duc 2012 for reviews).
TDGs do not contain significant amounts of dark matter because it cannot
be captured by their weak gravitational potentials (Barnes \&
Hernquist 1992; Kroupa 1997; Bournaud 2010; {  Kroupa 2012},
Dabringhausen \& Kroupa 2013, Kroupa 2015).

It is generally { thought} that TDGs do not build a mass-metallicity
(MZ) relation because their metallicity is dominated by recycled
material coming from the parent (large) interacting galaxies.  This is
certainly true for recently formed TDGs, for which in-situ production
of metals is negligible compared to the metals coming from the
interacting galaxies.  Indeed, recently formed TDGs are most easily
found as MZ outliers (Duc \& Mirabel 1994; 1998; Weilbacher et al.
2003; Croxall et al. 2009; but see also Reverte et al.  2007).
However, it is commonly argued that the production rate of TDGs was
much higher in the past, due to the larger gas fractions and the
smaller relative distances between galaxies.  Although the fraction of
surviving TDGs is unknown (see Kroupa 1997; Okazaki \& Taniguchi 2000;
Bournaud \& Duc 2006; Kaviraj et al. 2012; Dabringhausen \& Kroupa
2013 for different and conflicting estimates), numerical simulations
show that these galaxies can survive the tidal field of the parent
galaxy and the internal feedback processes for many Gyr (Kroupa 1997;
Klessen \& Kroupa 1998; Recchi et al. 2007; Casas et al. 2012;
Ploeckinger et al. 2014, 2015).  Ancient TDGs are therefore born out
of relatively unpolluted material and the subsequent chemical
evolution is solely due to internal processes.  They have thus the
possibility to build a MZ relation as any other galaxy (see also
Kroupa 2015).

The aim of this paper is to {  determine} the MZ relation of ancient
TDGs (born out of low-metallicity gas) and to compare them with the
observations of {  a sample of dwarf galaxies in the Local Universe.}
We want to test whether these objects (or a fraction of these objects)
might have had a tidal origin, as their internal dynamical properties
(Kroupa 1997; Metz \& Kroupa 2007) and the distribution { 
  of satellite galaxies about their host galaxies} in vast
{  thin} disks of satellites and their phase-space
correlation suggests ({  Kroupa, Theis \& Boily 2005};
Pawlowski et al. 2012; Ibata et al. 2013; Hammer et al. 2013;
Pawlowski \& Kroupa 2014; Yang et al.  2014; {  Pawlowski et
  al. 2014; Ibata et al. 2014a, 2014b, 2015}).  {  A recent paper
  (Collins et al. 2015) shows that the metallicity of dwarf galaxies
  belonging to the rotating disk of satellites surrounding Andromeda
  is comparable to the metallicity of off-plane satellites.  If the
  on-plane satellites are of tidal origin, one would expect instead
  large differences.  It is our aim to show that the mild difference
  can be explained if the on-plane dwarf galaxies are TDGs formed
  during the very early phases of the evolution of Andromeda and thus
  out of a very mildly polluted interstellar medium.}  At the same
time, models of younger TDGs, born out of polluted gas, will be
compared with observed TDGs.

In a companion paper (Recchi \& Kroupa 2015, hereafter Paper I) we
have shown how to calculate the metallicity (of stars and gas) of
model galaxies based on the {  integrated galactic IMF (IGIMF)}
theory (Kroupa \& Weidner 2003; Kroupa et al. 2013).  We have seen
that the IGIMF theory naturally produces a mass-metallicity (MZ)
relation, as low-mass galaxies have on average steep massive-star IMF
slopes and thus the metal production rate is reduced.  We have also
noticed that a good fit with the observed MZ relation in the stars of
local group dwarf galaxies (based on the Lee et al. 2006 data) can be
reproduced if one assumes that low-mass galaxies can expel gas and
metals through modest galactic winds.  Galactic winds are less likely
in large galaxies because of the deeper potential wells (Recchi \&
Hensler 2013).

In this paper we apply the methodologies of Paper I to the study of
the MZ relation in TDGs {  in order to test whether models of old
  TDG, born out of unpolluted or modetately polluted gas, can build a
  MZ relation.  We will apply the same models to models of younger
  TDGs, born out of more significantly polluted gas, in order to test
  whether the metallicity of observationally identified TDGs can be
  reproduced.}

The outline of this paper is as follows.  In Sect. \ref{sec:model} we
briefly recall the adopted methodology, extensively described in Paper
I.  In Sect. \ref{sec:results} the results of this investigation are
presented.  Finally, in Sect. \ref{sec:disc} the main results are
discussed and some conclusions are drawn.

\section{The method}
\label{sec:model}
We work in the framework of the IGIMF theory, according to which dwarf
galaxies, characterized by small levels of star formation rates
(SFRs), can produce only a small relative number of massive stars and,
hence, the galaxy-wide IMF will be biased toward low-mass stars.  In
particular, we adopt the detailed IGIMF prescriptions of Weidner et
al. (2013), which are able to reproduce a large range of observed
galactic properties.  We neglect the 'axiom $vii$' of Weidner et
al. (2013), namely we do not consider variations in the slope of the
mass distribution of star clusters.  We have shown in Paper I that
this assumption affects negligibly the results of our numerical
calculations.

{  A single high-resolution simulation of the birth and
  chemo-dynamical evolution of a TDG using hydrodynamical simulations
  with stellar feedback (Pl\"ockinger et al. 2014, 2015) is
  computationally very demanding, and a comprehensive survey of many
  models cannot be performed with available computers. Simplified
  models, which contain the essential physics, ar therefore
  unavoidable if a survey of the chemical and stellar-population
  properties of TDGs is to be made.  }  We assume the so called
'simple model' of chemical evolution: a one-zone model in which ejecta
from dying stars instantaneously mix with the surrounding gas.
Moreover, the instantaneous recycling approximation is adopted,
according to which the lifetime of stars with masses larger than
1~M$_\odot$ can be considered negligible.  This approximation is good
for $\alpha$-elements, produced on short timescales by massive stars,
but it is not good for elements produced on longer timescales by
intermediate-mass stars and Type Ia supernovae.  An outflow rate,
proportional to the SFR and dependent on the initial gaseous mass of
the model galaxy is assumed.  In particular, it is assumed that
low-mass galaxies, due to their shallower potential wells, develop
galactic winds and lose freshly produced heavy elements more easily
(Marin 2005; Recchi \& Hensler 2013).  {  An infall rate,
  proportional to the star formation rate, is assumed, too.}  The
solution of the simple model in the presence of {  infall and}
outflows is:
\begin{align}
  &M_g(t)=M_g(0)\exp{\left[(\Lambda-\lambda-1)s\int_0^t
      \left[1-R(\tau)\right]d\tau\right]},
  \notag\\
  &Z(t)=Z_0+\frac{\int_0^t[1-R(\tau)]s[\Lambda Z_A + y_Z(\tau)]
    I(\tau)d\tau}{I(t)}, \notag\\
  &I(t)=\exp{\left[-\int_0^t[1-R(\tau)]s[\lambda(\gamma-1)+\Lambda]
      d\tau. \right]}
  \label{eq:sol2}
\end{align}
\noindent
where $M_g$ is the gas mass, $Z$ is the metallicity, $R$ is the
fraction of a stellar population not locked into long-living (dark)
remnants, $y_{\rm Z}$ represents the ratio between the mass of heavy
elements ejected by a stellar generation and the mass locked up in
remnants, {  and $\gamma$ is the metallicity enhancement
  factor in the galactic wind (see Matteucci 2001; Recchi et al. 2008;
  Recchi 2014; Paper I for more details).  The SFR $\psi$ is
  parametrised as $\psi = s \times M_g$ with the star formation
  efficiency $s$ (i.e. the inverse of the gas-consumption time-scale)
  taken to be equal to 0.3 Gyr$^{-1}$ (c.f. Pflamm-Altenburg \& Kroupa
  2009).  The outflow rate is set by $\lambda (1-R) \Psi$ and the
  infall rate is determined with $\Lambda (1-R) \Psi$.}  {  As in
  Paper I, we assume $\Lambda=0.5$ and a variable $\lambda$, according
  to eq. 17 there.}  Finally, $Z_i$ is the initial metallicity { 
  and $Z_A$ is the metallicity of the infalling gas, which we assume
  to be equal to the initial metallicity $Z_i$}.  Adopting an
universal IMF, the quantities $R$ and $y_Z$ are constant and the
integrals appearing above can be explicitly calculated.  In the case
of the IGIMF, $R$ and $y_Z$ depend on the galactic evolution because
the SFR and the metallicity affect the IMF, which in turn affects the
calculation of these two quantities.  The above expression for $Z(t)$
is thus an implicit equation that must be solved iteratively.

The average metallicity $Z_*$ of the stars in a model galaxy is
calculated according to the equation
\begin{equation}
Z_*=\frac{\int_0^t Z (t) \psi (t)dt}{\int_0^t \psi(t)dt},
\label{eq:zstar}
\end{equation}
\noindent
where $\psi$ is the SFR.  This expression represents thus the
mass-weighted average of the metallicities of all the stellar
populations ever born in the galaxy (see Pagel 1997).

\section{Results}
\label{sec:results}
\subsection{Starless initial conditions}
\label{subs:starless}
At variance with Paper I, in which we considered only unpolluted
models of galaxies (i.e. we assumed $Z_i=0$ in Eq. \ref{eq:sol2}), we
consider here different levels of pre-enrichment.  In particular, in
what follows we consider mild pre-enrichments ($Z_i=10^{-3}$ and
$10^{-2}$ Z$_\odot$) and large pre-enrichments ($Z_i=0.1$ and $0.5$
Z$_\odot$).  In compliance with the results of Paper I, we consider
for the moment starless initial configurations, i.e. we assume that
the TDGs form out of the gaseous component of a tidal arm.  We will
relax this assumption in Sect. \ref{subs:stari}.  Notice that the
effect of $Z_i$ on the solution is not linear as Eq. \ref{eq:sol2}
might suggest because, according to the adopted IGIMF recipes, a
different initial metallicity changes also the initial IMF, and a
changing metallicity leads to a changing IGIMF.  We wish to consider
model solutions applicable to very old but also to young TDGs.  We
will compare the abundances of very old TDGs with the ones {  in a
  sample of dwarf galaxies in the local volume, including} dwarf
satellites orbiting around the Milky Way and Andromeda.

\begin{figure}
\resizebox{8.5cm}{!}{\includegraphics[angle=270]{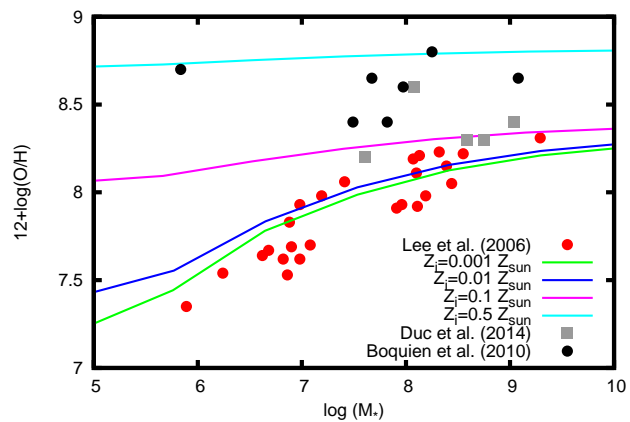}}   
\caption{The MZ relation obtained by means of the simple model of
  chemical evolution within the IGIMF theory, with different values of
  the initial metallicity $Z_i$ (see Eq. \ref{eq:sol2}).  Here, we
  compare the gas-phase abundance of the model galaxies with
  observations of {  dwarf galaxies in the Local Universe (from Lee
    et al. 2006; red circles) and of young} TDGs (from Boquien et al.
  2010 - {  black} circles; Duc et al. 2014 - { 
  grey} squares).  Notice that the $x$-axis indicates the final
stellar mass of the model galaxies, although the comparison focuses on
gas-phase abundances.  {  Notice also that the lower two curves
  ($Z_i=10^{-3}$ and $10^{-2}$ Z$_\odot$) correspond to old TDGs
  and evolve for a longer time (see text for details).}}
\label{fig:MZTDG}
\end{figure}

{  Notice that the dwarf satellites closest to the Milky Way and
  Adromeda are usually dwarf Spheroidals or dwarf Ellipticals.  For
  these galaxies, gas-phase abundances, obviously, can not be
  determined, and one resorts on stellar abundances (usually expressed
  as [Fe/H], as the iron composition of stars is easy to determine) to
  estimate the global metallicity of the galaxy.  Many observations
  are available concerning the average [Fe/H] in dwarf satellites and
  the resulting MZ relation (see e.g. Kirby et al. 2013).  In
  principle, these observations can be compared with our results,
  obtained by means of Eq.  \ref{eq:zstar}.  However, in Paper I we
  showed that the instantaneous recycling approximation can not be
  applied to iron, which is mainly produced on timescales longer than
  50 Myr (Matteucci \& Recchi 2001).  In paper I we thus compared our
  results with the sample of Lee et al. (2006).  This includes the
  gas-phase oxygen abundances of many gas-rich dwarf galaxies of the
  Local Group, as well as other dwarf galaxies of the Local Universe,
  not belonging to the Local Group.  Notice also that,} in the case of
younger TDGs, more data are available on the gas-phase abundances,
therefore we will compare the oxygen gas-phase abundance with
available observations in TDGs (taken from Boquien et al. 2010 and Duc
et al. 2014).  {  In order to have consistent datasets to compare
  with our results, and in order to be consistent with Paper I, we
  will consider only the gas-phase oxygen abundances of galaxies, and
  we will compare the results of models with low pre-enrichments with
  the observations of Lee et al. (2006), and the ones with high
  pre-enrichments with the results of Boquien et al. (2010); Duc et
  al. (2014).}  We need to consider also different evolutionary times
for the evolution of young and old TDGs.  We assume therefore an age
of 12 Gyr (the same evolutionary time considered in Paper I) for the
old TDGs and of 3 Gyr (see Duc et al.  2014) for the younger ones.
{  In particular, we assume here that the models with low
  pre-enrichment ($Z_i=10^{-3}$ and $10^{-2}$ Z$_\odot$) are old TDGs,
  evolving for 12 Gyr, whereas models with higher pre-enrichment
  ($Z_i=0.1$ and $0.5$ Z$_\odot$) are younger and evolve for 3 Gyr.}
The resulting theoretically derived MZ relations are shown in Fig.
\ref{fig:MZTDG}.

We can see from Fig. \ref{fig:MZTDG} that the models reproduce very
well the observations of {  Lee et al. (2006)} for mild levels of
pre-enrichment ($Z_i=10^{-3}$ or $10^{-2}$ Z$_\odot$) and that indeed
these results do not differ much from the ones with $Z_i=0$ presented
in Paper I.  On the other hand, larger values of $Z_i$ (up to $Z_i$
=0.5 Z$_\odot$) are required in order to fit the gas-phase abundances
of younger TDGs.

\subsection{TDGs versus dark matter dominated dwarf galaxies}
\label{subs:contrast}
\begin{figure}
\resizebox{8.5cm}{!}{\includegraphics[angle=270]{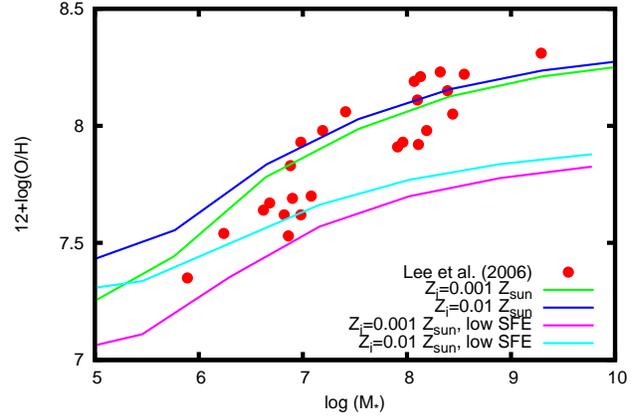}}   
\caption{The effect of different star formation efficiencies.  The
  curves $Z_{i}=0.001 Z_\odot$ and $Z_{i}=0.01 Z_\odot$ are the same
  as in Fig. \ref{fig:MZTDG}.  The other two curves have been obtained
  by halving the star formation efficiency $s$.  As in Fig.
  \ref{fig:MZTDG}, red circles correspond to the Lee et al. (2006)
  sample of galaxies}
\label{fig:lowSFE}
\end{figure}

{ 

Since we have shown that old TDGs do not stand
	out in a MZ relation as young TDGs do, the question
	arises on how we can possibly distinguish old
	TDGs from galaxies of similar sizes and ages but
	not of tidal origin.
	
	Assuming the gas infall for TDGs originates preferentially from the 
	tidal debris of the same galaxy interaction process, 
	not only the initial pre-enrichment, but also the infall parameters 
	differ from the DM-dominated dwarf galaxy case. 
	The gaseous material within the tidal arm is naturally close to the 
	TDG in phase-space, and can therefore be captured easily, 
	which could increase the infall rate. 
	
	However, in the framework of the simple models of
	chemical evolution we can not fix the infall rates, we
	can only fix the ratio between the infall rates and
	the SFR (the parameter $\Lambda$ introduced above). This
	scenario can be properly addressed only by means
	of more detailed but also computationally very demanding
	chemo-dynamical simulations, such as pioneered
	by Ploeckinger et al. (2014, 2015).
	
	In addition to a potentially increased infall rate, the infalling 
	tidal debris is homogeneously pre-enriched ($Z_A=Z_i$). 
	Tidal tails do not show a steep metallicity gradient as it is 
	typical for unperturbed
	late-type galaxies (Bresolin et al., 2009). Their metallicity 
	distribution is homogenised along the
	tidal arm (Kewley et al., 2010), possibly by radial gas mixing 
	during the galactic collision (Rupke
	et al., 2010). Contrary to that, the infall into
	DM-dominated dwarf galaxies is assumed to be of primordial 
	abundance ($Z_A = 0$). 
	
	If long-living TDGs form continuously throughout the history 
	of the Universe and therefore
	with a range of initial metallicities, one could expect that 
	they do not follow a tight MZ-relation
	but cover a large region in MZ diagram. The results of the 
	chemical evolution model presented in Fig. 1 
	explain why this is not the case. For low initial metallicities
	 ($<Z_i \approx 0.01\,\mathrm{Z}_{\odot}$), the theoretical
	MZ-relations are very similar to each other and the results are 
	almost indistinguishable whether the 
	infall has primordial abundance $Z_A = 0$ (Paper I) or the initial 
	metallicity $Z_A = Z_i$ (Fig. 1).
	Therefore, old TDGs that formed with a range of low initial 
	metallicities fall on the same MZ-relation. 
	
	Increasing the metal pre-enrichment of the tidal debris affects 
	both the initial metallicity of the TDG, and
	the metallicity of the infalling gas. This leads to an accelerated 
	evolution of the MZ-relation for 
	higher initial metallicities. Note the large difference between 
	the models for $Z_i = 0.1\,\mathrm{Z}_{\odot}$ and 
	$Z_i = 0.5\,\mathrm{Z}_{\odot}$ already after 3 Gyr (Fig. 1). 
	
	This accelerated evolution can therefore explain the gap between 
	the MZ-relation for old TDGs and position of the very young TDGs 
	in the MZ-diagram. {  Furthermore, if most of the
          satellite galaxies were born as TDGs at a high red shift,
          with the production rate of TDGs diminishing with cosmic
          time due to a decreased galaxy--galaxy encounter rate, then
          this would enhance the gap even further.}

        Another difference between TDGs and DM-dominated dwarf
        galaxies can be their star formation efficiencies, and this
        effect can be tested with our chemical evolution model. As an
        example, it is known for a long time that isolated galaxies
        have lower star formation efficiencies $s$ (i.e. longer
        gas-consumption time-scales, Pflamm-Altenburg \& Kroupa 2009)
        than interacting galaxies (e.g. Sanders \& Mirabel 1985,
        Solomon \& Sage 1988). If $s$ depends on the environment also
        on dwarf galaxy scales, the star formation efficiencies for
        DM-dominated dwarf galaxies and TDGs could be different. We
        can thus analyse what happens to the MZ relation if we use a
        different star formation efficiency. Fig.~\ref{fig:lowSFE}
        shows a comparison for the low pre-enrichment cases between
        our fiducial model with s = 0.3 Gyr$^{-1}$ and chemical
        evolution models where $s$ is decreased by a factor of 2
        (labelled with ``low SFE" in Fig.~\ref{fig:lowSFE}).  These
        models attain lower metallicities, mainly due to the lower
        SFRs (recall that, according to the IGIMF prescriptions, low
        SFRs correspond to an lower number of SNeII relative to an IMF
        description without a truncation at the massive star end).
   
        As seen in Fig.~\ref{fig:lowSFE}, while the low SFE lines
        match the observations of Lee et al. (2006) for stellar masses
        below $10^7\,\mathrm{M}_{\odot}$, the models underestimate the
        metallicities of systems with higher stellar masses. Detailed
        observational and numerical studies are required to constrain
        the star formation efficiency for old dwarf galaxies as well
        as for young TDGs, to further improve the accuracy of the
        chemical evolution models. }

\subsection{Initial conditions made of gas and stars}
\label{subs:stari}

We now assume that the initial TDG model contains gas and stars, both
recycled from the larger interacting galaxies.  In particular, we
assume that the initial configuration is characterized by an initial
stellar mass $M_{*, 0}$ which is equal to $\delta$ times the initial
gaseous mass $M_{g, 0}$.  These stars might come directly from the
parent galaxy, but they might also be the result of a tidally induced
burst of star formation.  According to the simple model assumptions
(see Paper I for a summary), a fraction of these stars die
instantaneously and restore $RM_{*, 0}=R \, \delta \, M_{g, 0}$ solar
masses of gas into the interstellar medium (ISM).  The initial gaseous
mass we have to consider in our models is thus $M_{g, 0}+R \, \delta
\, M_{g, 0}=M_{g, 0}(1+R\delta)$.  This pre-existing stellar
population releases also metals into the ISM.  The mass in metals at
the initial time is given by
\begin{equation}
M_Z(0)=M_{g, 0} \, Z_i + (1-R) \, M_{*, 0} \, y_Z.
\end{equation}
The first term represents the contribution due to the gas initially
present in the model galaxy, whereas the second term is the amount of
heavy elements instantaneously produced by a population of stars with
mass $M_{*, 0}$.  The initial metallicity of our model galaxy is thus:
\begin{equation}
Z(0)=\frac{M_Z(0)}{M_g(0)}=\frac{Z_i+(1-R) \, \delta \, y_Z}{1+\delta \, R}.
\end{equation}
\noindent
For small values of $\delta$, this initial metallicity correctly tends
to $Z_i$, but for larger values of $\delta$, the metallicity produced
by the initially present stars plays a prominent role and the initial
metallicity $Z(0)$ is less dependent on $Z_i$, i.e. the pristine
generation of stars injects so many metals into the ISM that the role
of the intial gaseous metallicity $Z_i$ becomes negligible.  Only the
initial SFR remains to be determined, as this SFR determines the
initial values of $R$ and $y_Z$ according to the IGIMF prescriptions.
We assume for simplicity that this initial SFR is given by $M_{*,
  0}/9$ M$_\odot$ Gyr $^{-1}$, i.e. it is the SFR required to build
$M_{*, 0}$ solar masses of stars in 9 Gyr at a constant rate (recall
that we are assuming that the TDG models evolve {  for} 3 Gyr
{  thereafter} in isolation).

\begin{figure*}
\resizebox{15cm}{!}{\includegraphics[angle=270]{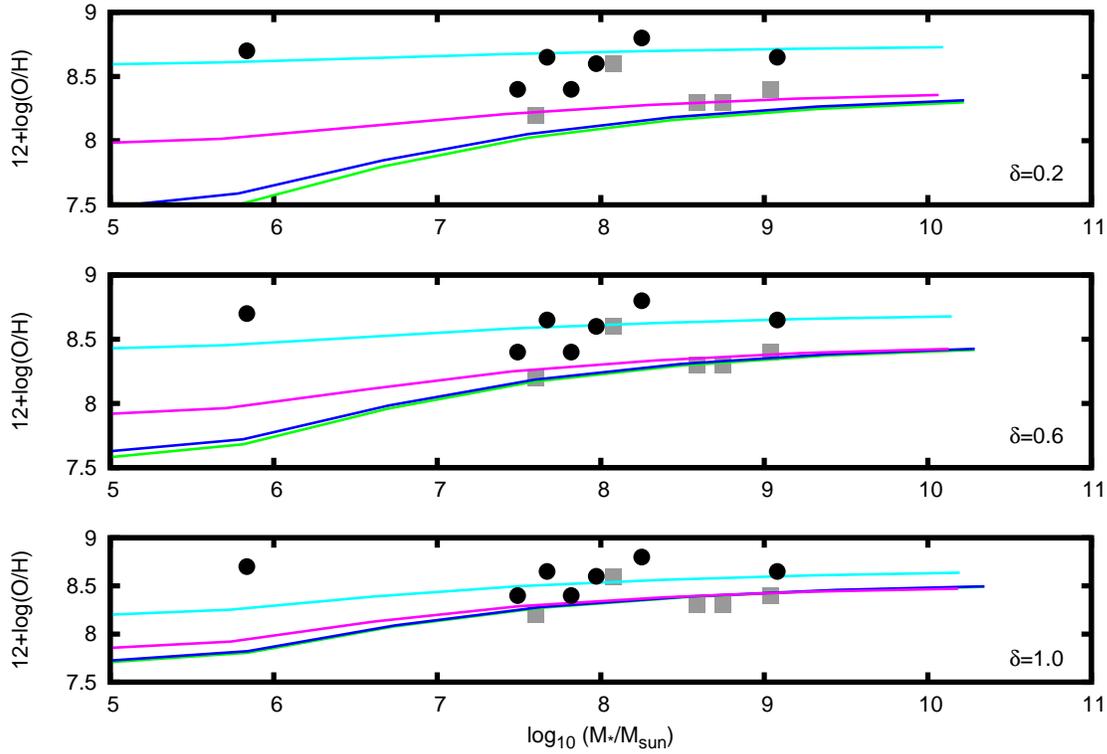}}
\caption{As in Fig. \ref{fig:MZTDG}, but for initial configurations
  containing already a population of stars.  The parameter $\delta$
  (indicated in each panel) represents the mass ratio between the
  initially present stellar component and the gaseous component, { 
    i.e. $\delta=\frac{M_{*, 0}}{M_{g, 0}}$}.  Color coding and
  symbols are as in Fig. \ref{fig:MZTDG}. }
\label{fig:MZinstars}
\end{figure*}

Fig. \ref{fig:MZinstars} shows the results of our calculations for
$\delta=0.2$, $\delta=0.6$ and $\delta=1.0$.  As predicted, the models
with $\delta=0.2$ differ very little compared to the models shown in
Fig. \ref{fig:MZTDG} (which correspond to $\delta=0$).  The agreement
with observations can still be considered satisfactory.  For larger
values of $\delta$, the various tracks get close to each other.  Once
again, this is due to the fact that the initial metallicity is set by
the instantaneous recycling of metals from the pristine stellar
population and $Z_i$ plays a less important role.

\section{Discussion and conclusions}
\label{sec:disc}

Many papers in the recent literature focus on the inconsistencies
between the results of cosmological hydrodynamical simulations and the
properties of dwarf galaxy satellites of the Milky Way (Klypin et
al. 1999; Moore et al.  1999; Kroupa et al. 2005; {  Kroupa
  et al. 2010}; Boylan-Kolchin et al. 2012 among others; see Kroupa
2012 for a review).  In particular, the spatial distribution of
satellites orbiting the Milky Way and Andromeda and their phase-space
correlations suggest {  rather strongly} that these objects
might be of tidal origin (Metz et al. 2007; Pawlowski et al. 2012).
Although many other hypotheses have been put forward to explain the
spatial distribution of dwarf satellites {  (e.g. Bahl \& Baumgardt
  2014)}, a detailed analysis reveals that their properties and
spatial arrangement fail to be accounted for if they were
dark-matter-dominated substructures (Ibata et al. 2014b; Pawlowski et
al. 2014; Kroupa 2015; Ibata et al. 2015).

If the dwarf satellites of the Milky Way and Andromeda (or a fraction
of them) are of tidal origin, one might expect their metallicity to be
very large, comparable to the metallicity of the external parts of
disks of large spiral galaxies, i.e. a few tenths of Z$_\odot$.
Clearly, dwarf satellites of the Milky Way and of Andromeda do not
attain such large metallicities.  On average, the metallicities are
very low, and there is a clear correlation between the stellar mass
and the stellar metallicity, although this correlation seems to
flatten out at low masses.  Indeed, recently formed TDGs, still
embedded in the parent tidal arm, have metallicities much larger than
their stellar mass would suggest (Duc \& Mirabel 1998; Weilbacher et
al. 2003), indicative of the fact that they are born out of recycled
material, processed by a very large galaxy.

It {  has been suggested} that the dwarf satellites of the Milky Way
can have been formed many Gyr ago as TDGs, during the early phases of
the evolution of the Milky Way and Andromeda (Kroupa et al. 2005; see
also Fouquet et al.  2012; Hammer et al. 2013; Yang et al. 2014; and
notably Zhao et al.  2013).  In this way, their initial metallicities
could have been very low and their initial gas fractions very high.
The subsequent chemical evolution of these objects is no longer
directly affected by the parent galaxies (only the tidal field might
have played a role) and follows the usual routes of isolated galaxies,
i.e. it is mainly affected by astration {  (the galactic cycle of
  chemical elements, due to ejection from dying stars and recycling
  through new generations of stars)} and, in the case of very low-mass
galaxies, galactic outflows.  On the other hand, younger TDGs, born in
the last few Gyr, are born out of material with a much higher
metallicity, thus ``endogenous'' processes (star formation, feedback,
galactic flows) play a minor role for the chemical evolution.

In this paper we have shown that chemical evolution models can well
reproduce the correlation between stellar mass and metallicity of both
young TDGs and { dwarf galaxies in the Local Group.}  The framework
here is the so-called IGIMF theory (Kroupa et al. 2013), according to
which small galaxies, characterized by mild levels of SFR, produce
very few massive stars and, hence, the nucleosynthesis of heavy
elements is severely limited compared to galaxies with higher masses.
According to our models, if the dwarf satellites are of tidal origin,
they must have been born out of metal-poor material and they must have
evolved for many Gyr.  If the levels of pre-enrichment range between
0.01 and 0.001 Z$_\odot$, then there is a very good fit between model
predictions and observations.  On the other hand, if the initial
metallicity is larger than 0.01 Z$_\odot$, and if the model galaxy
evolves in isolation for 3 Gyr, we are able to reproduce well the
observed gas-phase abundances in young TDGs (Boquien et al. 2010; Duc
et al. 2014).
If we consider initial conditions in which the mass of the stellar
component is a large fraction of the total baryonic mass, { 
  then} in this case, this pre-existing stellar population is able to
rise instantaneously the metallicity of the model galaxy and the
results are much less dependent on the initial metallicity.  The
metallicity of the gas of many of the observed real TDGs can
{  therefore} also be accounted for as having formed
containing a very significant pre-existing captured stellar component
from the host galaxy.

{  An interesting possible implication of this work is also
  that old TDGs (as suggested by the phase-space correlated satellite
  galaxies, Kroupa et al. 2005; Pawlowski et al. 2014; Ibata et
  al. 2013, 2014a, 2014b) and normal or primordial dwarf galaxies
  cannot be readily distinguished at a fundamental level (Kroupa 2015;
  Collins et al. 2015). This is one aspect of the failure of the dual
  dwarf galaxy theorem in a standard dark-matter based cosmological
  model, with the corresponding possible implications for
  gravitational physics (Kroupa 2012; 2015).  }

Implicit in our working hypothesis is the idea that TDGs (or a
substantial fraction of them) can survive the tidal torques
of the parent interacting galaxies and the internal feedback
processes for many Gyr.  Although the resilience of TDGs {  has been
  shown} in our previous numerical investigations (Kroupa 1997;
{  Klessen \& Kroupa 1998}; Casas et al.  2012; Recchi et
al. 2007; Ploeckinger et al. 2014, 2015), long-lasting {  (more than
  3 Gyr)} detailed numerical simulations of the chemo-dynamical
evolution of initially gas-rich TDGs are {  required in
  order to constrain this important problem further}.

\section*{Acknowledgements} 
{  We thank M. Drinkwater for {  very} useful and
  constructive {  comments and} M.  Pawlowski for very
  useful suggestions.  Exchanges of information and ideas with
  P.A. Duc are also greatly acknowledged. {  SP acknowledges 
  	support from the European Research Council
  under the European Union's Seventh Framework Programme
  (FP7/2007-2013)/ERC Grant agreement 278594-GasAroundGalaxies.}

\end{document}